\documentclass[nofootinbib,amssymb,floatfix,preprintnumbers,preprint]{revtex4}
\usepackage{graphicx}
\usepackage{epsfig}
\usepackage{dcolumn}
\usepackage{bm}
\usepackage{multirow}

\newcommand{\comment}[1]{}
\newcommand{\newc}{\newcommand}

\def\thetab0{\theta_{B_0}}

\def\r2{\sqrt 2}
\def\beq{\begin{equation}}
\def\eeq{\end{equation}}
\def\bea{\begin{eqnarray}}
\def\eea{\end{eqnarray}}

\def\wt{\widetilde}
\def\sinW2{\sin^2\theta_W}

\def\mz2{M_{z}^2}
\def\c2b{\cos 2\beta}

\def\mz{M_Z}

\def\sec2w{sec^2\theta_W}

\def\gmin2{(g-2)_\mu}

\catcode`\@=11 

\def\lsim{\mathrel{\mathpalette\@versim<}}
\def\gsim{\mathrel{\mathpalette\@versim>}}
\def\@versim#1#2{\vcenter{\offinterlineskipresembel
    \ialign{$\m@th#1\hfil##\hfil$\crcr#2\crcr\sim\crcr } }}

\newc{\ra}{\rightarrow}
\newc{\s}{\smallskip}
\newc{\non}{\noindent}

\def \chonep{{\wt\chi_1}^{+}}
\def \chonem{{\wt\chi_1^-}}
\def \chonep2{{\wt\chi_2^+}}
\def \chonem2{{\wt\chi_2^-}}

\def\mET{E_T \hspace{-1.2em}/\;\:}
\def\rpv{R_p \hspace{-0.92em}/\;\:}
\def\lfv{L_i \hspace{-0.9em}/\;\:}

\begin{document}


\title{Four lepton flavor violating signals at the LHC}

\author{Dilip Kumar Ghosh$^1$}%
\email{tpdkg@iacs.res.in}
\author{Probir Roy$^2$}%
\email{probir.roy@saha.ac.in}
\author{Sourov Roy$^1$}%
\email{tpsr@iacs.res.in}
\affiliation{$^1$Department of Theoretical Physics, 
Indian Association for the Cultivation of Science, 2A $\&$ 2B Raja
S.C. Mullick Road, Kolkata 700 032, India}
\affiliation{$^2$Saha Institute of Nuclear Physics, 1/AF Bidhannagar, Kolkata 700 064, India}
%
%
%
\date{\today}
\begin{abstract}
Some yet unknown dynamics is expected to be at work behind the flavor puzzles of the Standard Model. 
Speculations exist that this may manifest itself in significant strength at the terascale. One 
consequence may be lepton flavor violation with total lepton number conserved. Already observed in 
neutrino oscillation experiments, such a phenomenon may show up more prominently at TeV energies, thus 
signaling a completely new physics. Proposed flavor violating charged dilepton states have 
already been  studied with reference to the LHC. Here we study the production and detection at the LHC of
flavor violating charged quadrileptons which are shown to have certain 
advantages over dileptons in searching for lepton flavor violation. A classification of all six-fermionic operators, in 
the  chiral basis and contributing to such processes, is made and the corresponding cross section for each in 
14 TeV pp collisions is computed under the hypothesis of single operator dominance.  We further present the 
sensitivity reach of the new physics scale $\Lambda$ in terms of the integrated luminosity $\pounds$. 
\end{abstract}



\maketitle
\section{Introduction}
Flavor is an unresolved puzzle in the Standard Model (SM). On the one hand, the strengths of flavor 
changing weak neutral current transitions, observed so far, are as feeble as expected in the SM. On the other, we have no clear  
idea as to why charged fermion masses are hierarchically organized among the generations, nor why there is a mild 
hierarchy between the solar and atmospheric neutrino mass squared differences. Furthermore, there is no consensus on 
the reasons behind the rather small quark mixing angles and the relatively large neutrino mixing angles. A general 
expectation could be that of an underlying dynamics, yielding these features, to be revealed at some higher energy 
scale. New types of fields, which we generically call flavons \cite{Frogatt, Babu}, are likely to come into 
play there, mediating this dynamics and resulting in novel phenomena. Flavor issues in the SM invariably arise in 
connection with the mass matrices of elementary fermions in generation space, though they do affect observable gauge interactions 
among massive fermions. Those matrices, in turn, originate from Yukawa coupling matrices as a consequence of electroweak symmetry 
breaking. It may not be unreasonable, therefore, to expect \cite{Frogatt} the scale of flavor dynamics to be  within an 
order of magnitude of the electroweak scale and be accessible at the LHC. Any resultant runaway flavor-changing neutral 
current transition can be controled by linking \cite{Giudice} the flavor nonconserving couplings to the pattern of fermion mass matrices.

Our focus in this paper is on lepton flavor violation with total lepton number conserved. This has already been observed \cite{Fogli} in neutrino 
oscillation studies and is being probed \cite{Feldman, Goudelis, choubey} in yet unseen radiative and trileptonic decays of the $\mu$ and the $\tau$. 
If one were to extend the SM minimally just to include lightly massive and mixed Majorana or Dirac neutrinos (the ``$\nu$MSM''), that model would 
have such flavor nonconservation restricted largely to the neutrino sector. 
In particular, no observable charged lepton flavor violation would be expected to occur in multilepton production experiments 
at high energy accelerators. However, such an implication is precisely what we challenge here. We argue that flavor violation 
could occur at the LHC with charged multileptons and without hard neutrinos in the final state: a signal that would indicate a radically 
new physics. In this sense, our philosophy is fundamentally different from that of Minimal Flavor Violation \cite{isidori}. We specifically 
propose searching for charged and flavor violating quadrilepton signals in 14 TeV pp collisions at the LHC. Recent 
advances in developing strategies \cite{Chatrchyan} for multilepton searches at the LHC make this proposal very timely. 

The possibility of flavor violating charged multilepton signals was raised earlier \cite{Azuelos} in the  context of specific 
models. But we conduct here a more general study of final states with four charged and flavor violating leptons plus unobserved soft and forward 
hadrons at the LHC. The hard partonic subprocess that we have in mind is $q{\bar q}\rightarrow\ell_i^-\ell_j^-\ell_k^+\ell_n^+$. Here $\ell$ stands 
for a charged lepton with $i,j,k,n$ as flavor indices $e,\mu,\tau$, chosen in such a way that lepton flavor is not conserved.
A simpler subprocess would be $q{\bar q}\rightarrow\ell_i^-\ell_j^+$, $i\neq j$, 
giving rise to flavor violating charged dileptons in hadronic collisions. The case with $i\neq j=e,\mu$ is severely constrained by 
experimental \cite{pdg} limits on $\mu \rightarrow e$ conversion in nuclei with the corresponding new physics scale having been pushed beyond 
10 TeV which is  hence inaccessible to the LHC. However, a dileptonic final state with one $\tau$ is relatively unconstrained. A general study 
of possible $\mu\tau$ final states in pp collisions was carried out \cite{Black} some time ago. Our motivation for doing an analysis with charged 
quadrileptons instead is threefold, as explained below. 

Several factors motivate our choice. First, the SM background, coming mainly from decays of hadroproduced 4W's or 4$\tau$s, 
to our signal is much less than in the dilepton case; indeed, we shall argue that it can be made 
totally negligible here. Second, if there is a discrete symmetry or conservation law in the underlying model of flavor violation, the 
flavons \cite{Babu} -- or whichever new particles are responsible for the flavor violating phenomenon via their decays -- would be pair produced. 
In such a case four flavor violating leptons would make a very characteristic final state. Finally, unlike in the dilepton case, the production of 
a flavor violating quadrilepton final state consisting of only electrons and muons (which are easiest to detect and study), 
arising from the annihilation of a quark and an antiquark of the same flavor, is not particularly constrained by any other experiment, 
Mixed flavor violating four-lepton operators have been probed in searches \cite{Feldman} for trileptonic 
decays of the muon and the $\tau$ such as $\mu \rightarrow e {\bar e} e$, $\tau \rightarrow \mu {\bar \mu} \mu, \mu {\bar \mu} e, \mu {\bar e} e, 
e {\bar e} e$. However, we propose here a search for their production in quark-antiquark annihilation, probing six fermionic 
operators. Furthermore, configurations such as $e^\mp e^\mp \mu^\pm \mu^\pm$ or $\mu^\mp \mu^\mp \tau^\pm \tau^\pm$, which we can study in our
final state, are not accessible in those searched for decay processes.

As an illustration, we first consider the partonic subprocess $q^\alpha\bar q^\alpha\rightarrow\ell_i^-\ell_i^-\ell_j^+\ell_j^+ ~
(i\neq j = e,\mu,\tau)$ in 14 TeV pp collisions in an extension of the Minimal Supersymmetric Standard Model \cite{Drees} with R-parity violation, 
taken only in the leptonic sector, i.e. ${\rpv\lfv}$MSSM. Significant cross sections result at LHC14 for producing an $e^- e^- \mu^+ \mu^+$ final state. 
Thus motivated, we perform a general analysis of allowed operators with the lowest dimensionality, written in terms of SM fields in the chiral basis 
and contributing to flavor violating charged quadrilepton production in $q {\bar q}$ annihilation. We classify all independent six fermion operators 
(numbering twenty) that can contribute to the effective Lagrangian ${\cal L}_{eff}$. Assuming single operator dominance, the subprocess cross section 
is then computed for each. These are subsequently folded with parton distributions and appropriate cuts to compute the corresponding cross sections 
that are measurable in pp collisions at ${\sqrt s}$=14 TeV. Also discussed is the sensitivity reach on $\Lambda$ in terms of the integrated luminosity.

The rest of the paper is organized as follows. In Sec. II we compute the cross section for producing an $e^-e^-\mu^+\mu^+$ four lepton combination at 
$\sqrt{s_{pp}}$ = 14 TeV in our ${\rpv\lfv}$MSSM model. Sec. III is devoted to an enumeration of all twenty independent six fermionic operators 
contributing to the subprocess $q^\alpha\bar q^\alpha\rightarrow \ell_i^- \ell_j^- \ell_k^+\ell_n^+$ and to a discussion of the various permutation 
symmetry properties relating the corresponding production cross sections. Sec. IV contains numerical evaluations of the latter at 14 TeV $pp$ collisions 
at the LHC and of the associated sensitivity reach on the cutoff $\Lambda$. Finally, in Sec. V we summarize our conclusions.

\section{Flavor violating quadrilepton production in a model}
As mentioned earlier, the model we have in mind is ${\rpv\lfv}$MSSM. The lepton flavor violating part of its superpotential is given in 
usual superfield notation by \cite{Drees}
\bea
{\cal W}_{\rpv} = \frac{1}{2} \lambda_{[ij]k} {\hat L}_i {\hat L}_j {\bar {\hat E}}_k.
\eea
Here $i,j,k$ are lepton flavor indices and the square bracket around $i,j$ implies that the coupling strength $\lambda$ is antisymmetric between 
$i$ and $j$. There are consequent vertices between a sneutrino ${\tilde\nu}_k$ (as well as an antisneutrino ${\bar {\tilde \nu}}_k$) of flavor $k$ 
with charged leptons of flavors $i,j$ ($i \neq j$) so that the sneutrino/antisneutrino can have a transition into the pair $\ell_i^-\ell_j^+$. In 
the annihilation of $q$ and $\bar q$ of identical flavor, one can now have a diagram (Fig.1) with a virtual Z boson, propagating in the s-channel and 
creating an on-shell ${\tilde \nu}_k {\bar {\tilde \nu}}_k$ pair, each of which decays into $\ell_i^- \ell_j^+ (i \neq j)$. 
Taking into account the presence of two pairs of identical fermions in the final state, we can write the cross section for the process as
\cite{dawson-eichten-quigg}
\bea
&&\sigma(pp \rightarrow \ell_i^- \ell_j^+ \ell_i^- \ell_j^+ + X) 
=\frac{1}{4} \int dx_1 dx_2 \Sigma_q [q(x_1){\bar q}(x_2) + x_1 \leftrightarrow x_2] \times \nonumber \\
&&\frac{d{\hat \sigma}}{d {\hat t}}(q {\bar q} \rightarrow {\tilde \nu}_k {\bar {\tilde \nu}_k}) \times Br({\tilde \nu}_k \rightarrow 
\ell_i^- \ell_j^+) \times Br({\bar {\tilde \nu}_k} \rightarrow \ell_i^- \ell_j^+), 
\eea
where
\bea
\frac{d{\hat \sigma}}{d {\hat t}}(q {\bar q} \rightarrow {\tilde \nu} {\bar {\tilde \nu}})
= \frac{e^4 (c+t)^2}{96 \pi {\hat s}^2} ({\hat t} {\hat u} - m^4_{\tilde \nu}) 
(A_q^2 + B_q^2)|D_Z({\hat s})|^2. 
\eea
Here ${\hat s},{\hat t}, {\hat u}$ are Mandelstam variables for the
subprocess,  $D_Z({\hat s}) = 1/({\hat s} - M^2_Z)$, $A_q = -\frac{5}{12}t + \frac{1}{4}c$ for $q =u,c$ and $A_q = -\frac{1}{4}c + \frac{1}{12}t$ for 
$q =d,s,b$. $B_q = -\frac{1}{4}(c + t)$ for $q =u,c$ and $B_q = +\frac{1}{4}(c + t)$ for $q = d,s,b$ with $c = \cot\theta_W$, and 
$t = \tan\theta_W$, $\theta_W$ being the Weinberg angle. Taking $\lambda_{[ij]k}$ and $\lambda_{[ik]j}$ to be the same, we obtain the product 
of the two decay branching fractions to be 1/4. 

We obtain the signal cross section using the CTEQ6L parton distributions 
\cite{cteq6l} with the factorization scale $Q^2 = {\hat s}/4 $. 
We also impose the following set of criteria for our signal events:
\begin{eqnarray}
p_T^\ell > 10~{\rm GeV},~\mid \eta^\ell \mid < 2,~\Delta R_{\ell \ell} > 0.4,
\end{eqnarray}
where all these symbols have their usual meaning. Our signal events are 
free from any real missing transverse energy. However, in practice 
there would be a minimum resolution for missing $E_T$ in a 
hadron collider such as the LHC. To take this into account, we 
impose an additional requirement on our signal events:
\begin{eqnarray}
\mET  < 20~{\rm GeV}.
\end{eqnarray}
We show a plot (Fig.2) of the cross section for the process 
with $i=e,j=\mu,k=\tau$ as a 
function of the $\tau$ sneutrino mass for 14 TeV pp collisions. For the situation of interest in which the said mass varies from 100 GeV to 1 TeV, 
the cross section for the production of $e^- e^- \mu^+ \mu^+$ in 14 TeV pp collision goes monotonically down from about 6 fb to slightly 
below 10$^{-3}$ fb, cf. Fig.\ref{cross-section-snu-tau}.
\begin{figure}[htb]
\centering
\includegraphics[width=6.40cm]{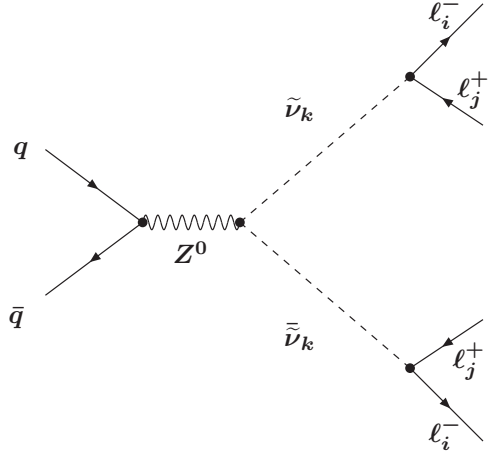}
\caption{Diagram for flavor violating charged quadrilepton production in the 
$\rpv {\lfv}$ MSSM model.}
\label{diagram-lfv}
\end{figure}

\begin{figure}[htb]
\centering
\vspace*{-0.5in}
\includegraphics[width=7.40cm]{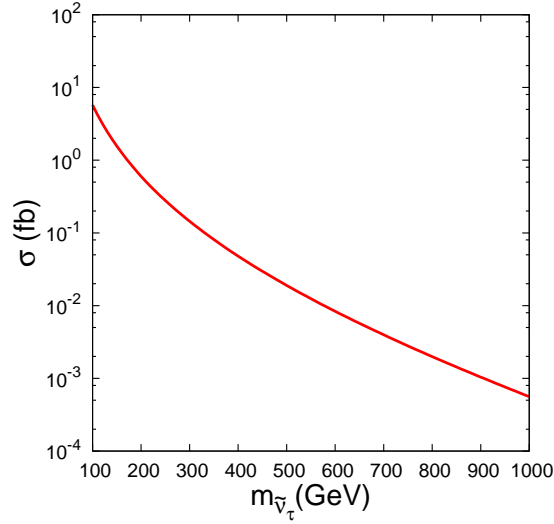}
\vspace*{-0.5in}
\caption{Cross section for $pp \rightarrow e^- e^- \mu^+ \mu^+ + X$ at 14 TeV 
in the ($\rpv {\lfv}$) MSSM model 
as a function of $m_{{\tilde \nu}_\tau}$.} 
\label{cross-section-snu-tau}
\end{figure}

\section{Operator description}

We treat the work of Ref.\cite{Black} on a final state with $\mu$,$\tau$ and soft/forward hadrons from pp collisions, 
probing a type of two-quark, two-lepton operators \cite{Carpentier}, as our starting point. Total baryon and lepton numbers were 
kept conserved here and an effective Lagrangian ${\cal L}_{eff}$ of mass dimension six for the partonic subprocess 
$q^\alpha{\bar q^\alpha} \rightarrow \ell_i^- \ell_j^+$ was taken to be 
\bea
{\cal L}^{ij}_{eff}={\mathbf \Sigma}_{p,\alpha} c^p_\alpha \Lambda_{p,\alpha}^{-2} (\bar\ell_i \Gamma_p \ell_j) (\bar q^\alpha \Gamma_p q^\alpha) 
+ h.c,
\label{han-effective}
\eea
with $i,j$ chosen as $\mu,\tau$ respectively. In Eq.\ref{han-effective}, 
$c^p_\alpha$ are dimensionless coefficients and   $\Lambda_{p,\alpha}$, 
presumed to be $\cal O$(TeV), is the scale of the new physics underlying the effective operator. The index $\alpha$, to be summed incoherently in the 
cross section, refers to the flavor of the annihilating quark-antiquark pair. Different flavors for the quark and the antiquark need not be 
considered because of restrictions from the non-observation \cite{pdg} of decays such as $B^0, D^0\rightarrow\mu\tau$. 
Furthermore, $\Gamma_p$ is a generic Dirac matrix: $\Gamma_p \in (1,\gamma_5,\gamma_\sigma, \gamma_\sigma\gamma_5)$.
Tensor products and leptoquark bilinears were not included; they can always be 
cast by Fierz transformations \cite{Buchmuller} into the form of 
Eq.\ref{han-effective}. 
Operators involving gauge  bosons \footnote{ 
This is on account of the assumption in Ref. \cite {Black} that the underlying 
fermionic couplings are like that of a strongly interacting gauge theory. If,
in contrast, they were very weak (like Yukawa couplings), operators with gauge
bosons could dominate, as happens with gluon-gluon fusion into the SM Higgs 
boson. We thank M. Hirsch for emphasizing this point.} generally yield 
contributions suppressed by 
${\cal O}(M_W/\Lambda)$ or ${\cal O}(m_{top}/\Lambda)$ 
terms and were ignored. The coefficients $c^p_\alpha$ were taken to be 
${\cal O}$(4$\pi)$ with an underlying strongly interacting gauge theory, 
characterized by a fine structure coupling strength of order unity, in mind. 

In writing a corresponding effective Lagrangian for our 
concerned partonic subprocess in terms of SM fields, we find it convenient 
to work in 
the chiral basis. This procedure has two advantages. (1) One can keep the electroweak symmetry properties of the operators as manifest. 
(2) Different operators differ in some chirality or other of one or more of the fermions, thereby contributing incoherently to the  subprocess 
with zero interference. Once overall baryon and lepton conservation are imposed, one can see that every operator always contains an odd number 
of left-chiral fermion doublet fields. $SU(2)_L$ invariance then requires that the relevant six chiral fermionic fields be accompanied 
multiplicatively by the SM Higgs doublet field H. Simply, the VEV of the neutral component of the latter can be taken to contribute to our 
${\cal L}_{eff}$. Since this component has only weak interactions, it is reasonable to further assume the multiplicative occurrence of the 
semiweak $SU(2)_L$ coupling constant from the underlying theory and posit the following form for the effective Lagrangian leading to our subprocess:
\bea
{\cal L}_{eff} = {\mathbf \Sigma}_{{p,r,s},{a,...e},\alpha}c^I(\Lambda_I)^{-6} M_W
                          \bar q^\alpha_a\Gamma_pq^\alpha_b\bar\ell_{ic}\Gamma_r\ell_{kd}{\bar \ell}_{je}\Gamma_s\ell_{nf}. \nonumber \\
\label{eq:lag_eff}
\eea
In (\ref{eq:lag_eff}), the subscripts $a,b,c,d,e,f$ are chiral indices (L or R, as applicable), $p,r,s$ are appropriately contracted Lorentz indices, 
$\alpha$ is a quark flavor index which needs to be summed incoherently in the cross section, $i,j,k,n$ are fixed lepton flavor indices chosen so 
as to ensure lepton flavor violation and $I$ is an all encompassing index $I\in(a,b,c,d,e;p,r,s,\alpha,i,j,k,n)$. The dimensionless coefficients 
$c^I$ are expected to be of order unity. For simplicity, we take them to be one for all $I$ and also drop the subscript $I$ on $\Lambda$ taking 
the cutoff to be universal for all the operators. We are now able to classify all six fermionic operators that can contribute to ${\cal L}_{eff}$. 
There turn out to be twenty such independent operators in the chiral basis. The cross section for the subprocess 
$q\bar q\rightarrow\ell_i^-\ell_j^-\ell_k^+\ell_n^+$ can be calculated for each one of these. Assuming single operator dominance, we can take 
the latter to be the cross section for the production process mediated by the concerned operator. 

Each independent operator, contributing to the product ${\bar q}^\alpha_a \Gamma_p q^\alpha_b {\bar\ell}_{ic} \Gamma_r \ell_{kd} {\bar \ell}_{je} 
\Gamma_s \ell_{nf}$, will be constructed now. Let us first make four remarks. (1) The notation $\ell_{iL},\ell_{jR}$ is used respectively for the 
$SU(2)_L$ doublet component and singlet charged lepton fields of flavor $i,j$. Because of $SU(2)_L$ invariance, one can also write similar operators 
replacing $\ell_{iL}$ by $\nu_i$, but we are not interested in those here. (2) All $\Gamma$ combinations with one, two or three tensor matrices 
$\sigma_{\mu\nu}$ can be reduced \cite{Buchmuller, Bailin} by Fierz transformations to scalar or vector products between chiral fields. (3) We are 
finally left with two sets of operators: Set I comprising  products of three scalar chiral fermion bilinears and Set II consisting of products of 
one scalar chiral fermion bilinear with two contracted vector chiral fermion bilinears. (4) In each member of the first set of operators, all three 
bilinears are disconnected from one another; this set of operators is called disconnected and designated D. Among the three bilinears of every 
member of the second set, two are connected by contraction while the third is disconnected from the rest; this set is called semi-disconnected 
and designated S. It is noteworthy that operators with all three fermion bilinears connected with one another by contraction are automatically 
absent from our compilation.

We list in Tables \ref{disconnected-operators} and \ref{Semi-disconnected-operators} the precise form of each operator in Sets I and II respectively. 
Each has superscripts $i,j,k,n$ for the observed lepton flavors and subscripts N$\alpha$, N=1,2... being a label for the 
\begin{table}[ht]
\scriptsize
\caption{\label{disconnected-operators}
Purely disconnected operators.
}
\begin{ruledtabular}
\begin{tabular}{ c  c }

     Label           &                                              Form \\ \hline \\

  $D^{ijkn}_{1\alpha}$   &          $\Lambda^{-6} M_W {\bar q}^\alpha_L q^\alpha_R {\bar \ell}_{iL} \ell_{kR} {\bar \ell}_{jL} \ell_{nR}$ \\ \\

  $D^{ijkn}_{2\alpha}$   &          $\Lambda^{-6} M_W {\bar q}^\alpha_R q^\alpha_L {\bar \ell}_{iL} \ell_{kR} {\bar \ell}_{jl} \ell_{nR}$ \\ \\

  $D^{ijkn}_{3\alpha}$   &          $\Lambda^{-6} M_W {\bar q}^\alpha_L q^\alpha_R {\bar \ell}_{iR} \ell_{kL} {\bar \ell}_{jR} \ell_{nl}$ \\ \\

  $D^{ijkn}_{4\alpha}$   &          $\Lambda^{-6} M_W {\bar q}^\alpha_R q^\alpha_L {\bar \ell}_{iR} \ell_{kL} {\bar \ell}_{jR} \ell_{nl}$ \\ \\
  
  $D^{ijkn}_{5\alpha}$   &          $\Lambda^{-6} M_W {\bar q}^\alpha_L q^\alpha_R {\bar \ell}_{iL} \ell_{kR} {\bar \ell}_{jR} \ell_{nl}$ \\ \\

  $D^{ijkn}_{6\alpha}$   &          $\Lambda^{-6} M_W {\bar q}^\alpha_R q^\alpha_L {\bar \ell}_{iL} \ell_{kR} {\bar \ell}_{jR} \ell_{nl}$
\end{tabular}
\end{ruledtabular}
\end{table}
\begin{table}[ht]
\scriptsize
\caption{\label{Semi-disconnected-operators}
Semi-disconnected operators.}
\begin{ruledtabular}
\begin{tabular}{ c  c }
     Label                  &                                        Form \\ \hline \\

$S^{ijkn}_{1\alpha}$ & $\Lambda^{-6} M_W {\bar q}^\alpha_L q^\alpha_R {\bar\ell}_{iL} \gamma_\rho \ell_{kL} {\bar \ell}_{jL} \gamma^\rho \ell_{nL}$ \\ \\

$S^{ijkn}_{2\alpha}$ & $\Lambda^{-6} M_W {\bar q}^\alpha_R q^\alpha_L {\bar \ell}_{iL} \gamma_\rho \ell_{kL} {\bar \ell}_{jL} \gamma^\rho \ell_{nL}$ \\ \\

$S^{ijkn}_{3\alpha}$ & $\Lambda^{-6} M_W {\bar q}^\alpha_L q^\alpha_R {\bar \ell}_{iR} \gamma_\rho \ell_{kR} {\bar \ell}_{jR} \gamma^\rho \ell_{nR}$ \\ \\

$S^{ijkn}_{4\alpha}$ & $\Lambda^{-6} M_W {\bar q}^\alpha_R q^\alpha_L {\bar \ell}_{iR} \gamma_\rho \ell_{kR} {\bar \ell}_{jR} \gamma^\rho \ell_{nR}$ \\ \\

$S^{ijkn}_{5\alpha}$ & $\Lambda^{-6} M_W {\bar q}^\alpha_L q^\alpha_R {\bar \ell}_{iL} \gamma_\rho \ell_{kL} {\bar \ell}_{jR} \gamma^\rho \ell_{nR}$ \\ \\

$S^{ijkn}_{6\alpha}$ & $\Lambda^{-6} M_W {\bar q}^\alpha_R q^\alpha_L {\bar \ell}_{iL} \gamma_\rho \ell_{kL} {\bar \ell}_{jR} \gamma^\rho \ell_{nR}$ \\ \\

$S^{ijkn}_{7\alpha}$ & $\Lambda^{-6} M_W {\bar q}^\alpha_L \gamma_\rho q^\alpha_L {\bar \ell}_{iL} \gamma_\rho \ell_{kL} {\bar \ell}_{jL} \ell_{nR}$ \\ \\

$S^{ijkn}_{8\alpha}$ & $\Lambda^{-6} M_W {\bar q}^\alpha_L \gamma_\rho q^\alpha_L {\bar \ell}_{iL} \gamma_\rho \ell_{kL} {\bar \ell}_{jR} \ell_{nL}$ \\ \\

$S^{ijkn}_{9\alpha}$ & $\Lambda^{-6} M_W {\bar q}^\alpha_R \gamma_\rho q^\alpha_R {\bar \ell}_{iR} \gamma_\rho \ell_{kR} {\bar \ell}_{jL} \ell_{nR}$ \\ \\

$S^{ijkn}_{10\alpha}$ & $\Lambda^{-6} M_W {\bar q}^\alpha_R \gamma_\rho q^\alpha_R {\bar \ell}_{iR} \gamma_\rho \ell_{kR} {\bar \ell}_{jR} \ell_{nL}$ \\  \\

$S^{ijkn}_{11\alpha}$ & $\Lambda^{-6} M_W {\bar q}^\alpha_L \gamma_\rho q^\alpha_L {\bar \ell}_{iR} \gamma_\rho \ell_{kR} {\bar \ell}_{jL} \ell_{nR}$ \\ \\

$S^{ijkn}_{12\alpha}$ & $\Lambda^{-6} M_W {\bar q}^\alpha_L \gamma_\rho q^\alpha_L {\bar \ell}_{iR} \gamma_\rho \ell_{kR} {\bar \ell}_{jR} \ell_{nL}$ \\ \\

$S^{ijkn}_{13\alpha}$ & $\Lambda^{-6} M_W {\bar q}^\alpha_R \gamma_\rho q^\alpha_R {\bar \ell}_{iL} \gamma_\rho \ell_{kL} {\bar \ell}_{jL} \ell_{nR}$ \\ \\

$S^{ijkn}_{14\alpha}$ & $\Lambda^{-6} M_W {\bar q}^\alpha_R \gamma_\rho q^\alpha_R {\bar \ell}_{iL} \gamma_\rho \ell_{kL} {\bar \ell}_{jR} \ell_{nL}$ 
\end{tabular}
\end{ruledtabular}
\end{table}
operator. We have not listed 
the hermitian conjugates of the operators since they lead to a different final state, namely $\ell_k^- \ell_n^- \ell_i^+ \ell_j^+$. 
However, that is recoverable from our final state by an ${i,j}\leftrightarrow{k,n}$ interchange. 
Given our choice of the value of unity for the coefficients $c^I$ and of universality for the cutoff scale $\Lambda$, such an interchange has become 
a symmetry of ${\cal L}_{eff}$ in consequence. Finally, since the lepton masses are neglected, the expressions for the square of the invariant matrix 
element and the cross section should become independent of lepton flavor labels.

Let us assign four momenta $p_1$ and $p_2$ respectively to the $q$ and $\bar q$, $k_1$ and $k_2$ respectively to $\ell_i^-$ and $\ell_j^-$, $q_1$ 
and $q_2$ respectively to $\ell_k^+$ and $\ell_n^+$ with $p_1+p_2=k_1+k_2+q_1+q_2$. The exprssions for the initial spin-averaged and final 
spin-summed square of the invariant matrix element (called $R$) for the different operators are now listed in Table \ref{expressions-R} in 
the approximation of treating all concerned fermions as massless. The consequent analytic expressions for quadriple differential cross sections 
can be simply derived from these. The combination of hermitian conjugation and ${i,j}\leftrightarrow{k,n}$ interchange yields pairs of operators 
which must lead to the same expressions for $R$. These pairs are $(D_1,D_4), ~(D_2,D_3), ~(D_5,D_6), ~(S_1,S_2), ~(S_3,S_4),$ $(S_5,S_6), 
~(S_7,S_8), ~(S_9,S_{10}), ~(S_{11},S_{12}), ~(S_{13},S_{14})$. Furthermore, in computing $R$, the contractions 
of the chiral fermionic fields (taken to be massless) are similar for each member of the purely disconnected set $(D_1,D_2,D_3,D_4)$. Therefore, 
they all lead to the same result, while $(D_5,D_6)$ yield something different. Turning to the second set of semi-disconnected operators, a similar 
argument applies separately to members of the sets $(S_1,S_2,S_3,S_4)$, $(S_5,S_6)$, $(S_7,S_8,S_9,S_{10})$ and $(S_{11},S_{12},S_{13},S_{14})$ since 
the chiralities L and R behave identically when they are not mixed. This means that the  operators $(S_1,S_2,S_3,S_4)$ all lead to an identical 
expression for $R$; the same can be said separately about $(S_5,S_6)$, $(S_7,S_8,S_9,S_{10})$ as well as  $(S_{11},S_{12},S_{13},S_{14})$. However, 
the four expressions for the said quantity emerging from the four sets of S-operators are different from one another. The exprssions for $R$ from 
the different operators are listed in Table \ref{expressions-R} in the approximation of treating all concerned fermions as massless. Comparing the 
two tables, we note that the expression for $(S_5,S_6)$ is just four times that for $(D_5,D_6)$. 
\begin{table}[ht]
\scriptsize
\caption{\label{expressions-R}
Expressions for $R$.
}
\begin{ruledtabular}
\begin{tabular}{ l  l }

Operators &  Expression for $R$ \\ \hline \\

$D_1,D_2,D_3,D_4$   &   $2M_W^2\Lambda^{-12}p_1.p_2 k_1.k_2 q_1.q_2$ \\ \\

$D_5,D_6$      &   $M_W^2\Lambda^{-12}p_1.p_2(k_1.q_1 k_2.q_2 + k_1.q_2 k_2.q_1)$ \\ \\

$S_1,S_2,S_3,S_4$  &   $8M_W^2\Lambda^{-12}p_1.p_2(3k_1.k_2 q_1.q_2 - k_1.q_1 k_2.q_2 - k_1.q_2 k_2.q_1)$ \\ \\

$S_5,S_6$  &  $4M_W^2\Lambda^{-12}p_1.p_2(k_1.q_1 k_2.q_2 + k_1.q_2 k_2.q_1)$ \\ \\

$S_7,S_8,S_9,S_{10}$    & $2 M_W^2\Lambda^{-12}[p_1.q_2(p_2.q_1 k_1.k_2 + p_2.k_2 k_1.q_1 + p_2.k_1 k_2.q_1)$ \\
                        &  $+ p_1.q_1(p_2.q_2 k_1.k_2 + p_2.k_2 k_1.q_2 + p_2.k_1 k_2.q_2)$ \\
                        &  $- p_1.k_2(p_2.q_2 k_1.q_1 + p_2.q_1 k_1.q_2)$ \\
                        &  $- p_1.k_1(p_2.q_2 k_2.q_1 + p_2.q_1 k_2.q_2)]$ \\ \\

$S_{11},S_{12},S_{13},S_{14}$ &   $2 M_W^2\Lambda^{-12}[p_1.k_2(p_2.q_2 k_1.q_1 + p_2.q_1 k_1.q_2 + p_2.k_1 q_1.q_2)$ \\
                              &    $+ p_1.k_1(p_2.q_2 k_2.q_1 + p_2.q_1 k_2.q_2 + p_2.k_2 q_1.q_2)$ \\
                              &   $- p_1.q_2(p_2.k_2 k_1.q_1 + p_2.k_1 k_2.q_1)$ \\
                              &   $- p_1.q_1(p_2.k_2 k_1.q_2 + p_2.k_1 k_2.q_2)]$.

\end{tabular}
\end{ruledtabular}
\end{table}

Coming to total cross sections, clearly the four effectively massless fermions in the final state develop a symmetry under the interchange of 
any two pairs upon phase space integration. This means that all six D-operators will lead to the same total cross section which is one fourth of the 
value of that of the set $(S_1,S_2,S_3,S_4,S_5,S_6)$, all members of which will yield the same total cross section. By the same token, all members 
of the set $(S_7,S_8,S_9,S_{10},S_{11},S_{12},S_{13},S_{14})$ will have an identical total cross section. Though this cross section is numerically 
close to that of the set $(D_1,D_2,D_3,D_4,D_5,D_6)$, it is in fact discernably different from the latter.

\section{Signals and sensitivity}
We first recount the signal of the process discussed in Ref.\cite{Black}. The partonic cross section, ignoring fermion mass terms, works out 
to be approximately $(4\pi)^2 {\hat s} \Lambda^{-2}$, $\sqrt{{\hat s}}$ being the partonic center-of-mass (CM) energy of the $q{\bar q}$ pair. 
In calculating measurable event rates, the authors of Ref.\cite{Black} used strong lepton $p_T$ and azimuthal cuts and kept the invariant $\mu\tau$ 
mass above 250 GeV to drastically reduce a sizable SM background from WW and Drell-Yan $\tau\tau$ production. A detailed analysis,
considering both leptonic and hadronic decays of the $\tau$, led to an estimated $\tau$ detection efficiency of $\sim$ 0.67. The 
sensitivity reach of $\Lambda$ at the LHC, with $\sqrt{s}$ = 14 TeV and an integrated luminosity of 100 fb$^{-1}$, was found to be in the 
6-21 TeV range for different operators. 
On the other hand, the validity of perturbative unitarity requires $\sqrt{\hat s} < {\cal O}(\Lambda)$. Since the maximal parton 
luminosity for the 14 TeV LHC run is expected \cite{Mandy} to be at $\sqrt{\hat s}$=$\sqrt{{x_1}_{max}{x_2}_{max}s}$ $\sim$
$\sqrt{0.3 \times 0.03s}$ $\sim$ 420 GeV, $x_{1,2}$ being the quark, antiquark partonic fraction, the statement on the $\Lambda$ reach 
was unitarity safe.

We come now to flavor violating quadrilepton production at the LHC. Charged multileptons, not conserving flavor but being accompanied by a 
large missing transverse energy ($\mET$ in short), cannot a priori qualify as genuine signals of lepton flavor violation. This is since 
neutrinos of compensating flavor could escape as part of the missing $E_T$. Indeed, a preliminary negative search with 35 pb$^{-1}$ of 
data for the final state $\mu^\mp e^\pm$ +  {$\mET$} has already been carried out by ATLAS \cite{Aad}. Such charged multileptons of unmatched 
flavors admit copious production, along with a large {$\mET$}, in a number of scenarios beyond the Standard Model (BSM) with or without 
lepton flavor violation. Some examples are models with new neutrino interactions \cite{delAguila}, those with heavy Majorana fermions \cite{Perez} 
and ones with supersymmetry \cite{Carquin}. In order to avoid confusion 
with such other BSM scenarios, we insist on no missing $E_T$ or $0\mET$ 
in the final state as a key signal criterion. 
In order to sharpen our leptonic signals and to avoid overlap 
with other production mechanisms, the absence of hard jets from the final 
state is required; only soft and forward hadrons and minijets from $\tau$'s are 
allowed. We therefore zero in on a $0\mET~0j\ell_i^-\ell_j^-\ell_k^+\ell_n^+$ signal where the four indices $i,j,k,n$ do not add up to 
flavor conservation. 

\begin{figure}[htb]
\centering
\vspace*{-0.5in}
\includegraphics[width=7.40cm]{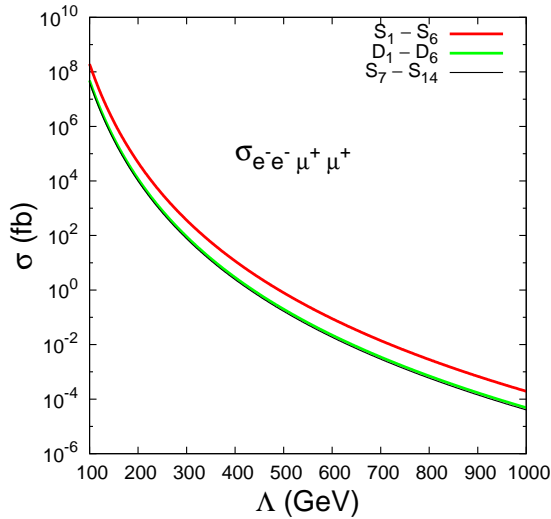}
\vspace*{-0.8in}
\caption{Cross section ($\sigma$) at 14 TeV 
with the $e^- e^- \mu^+ \mu^+$ final state as a function of the cut-off 
scale $\Lambda$ for different operators as shown.} 
\label{cross-section-dis}
\end{figure}

\begin{figure}[htb]
\centering
\includegraphics[width=7.40cm]{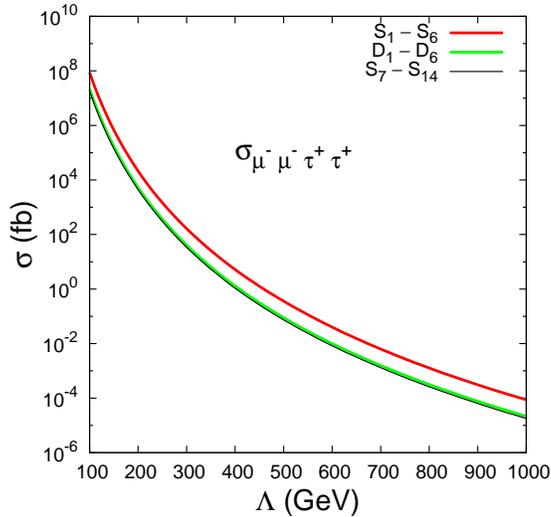}
\vspace*{-0.8in}
\caption{Similar plots as in Fig.\ref{cross-section-dis}, but for the $\mu^- \mu^- \tau^+ \tau^+$ final state.}
\label{cross-section-semi-dis1}
\end{figure}

We estimate our signal rates for 14 TeV pp collisions with same choice
of parton distribution, the factorization scale and selection criteria 
(Eqs.4 and 5) as discussed in Section II. 
Fig.\ref{cross-section-dis} shows plots of total cross sections separately 
for each set of operators for the 
$ee\bar\mu\bar\mu$ case. In relation to $\tau$'s, we follow Ref.\cite{Black}. Employing both leptonic (~0.35 BR) and hadronic (~0.65 BR) channels 
as well as appropriate lepton ($p_T$, $\Delta R$)and $\tau$-jet cuts, a net detection efficiency of ~0.67 is used. 
Fig.\ref{cross-section-semi-dis1} shows the corresponding plots in the $\mu \mu {\bar \tau} {\bar \tau}$ case. 
\begin{figure}[htb]
\centering
\includegraphics[width=7.40cm]{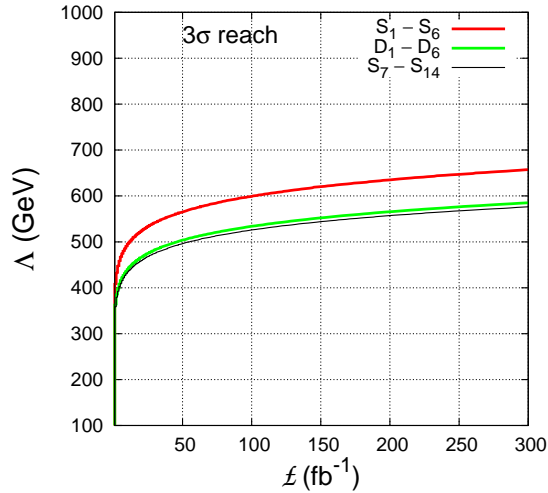}
\vspace*{-0.7in}
\caption{3$\sigma$ reach of the cut-off scale $\Lambda$ as a function of the total integrated luminosity ${\pounds} ({\rm fb}^{-1})$ 
for the various operators discussed in the text 
with the $e^-e^-\mu^+ \mu^+$ final state.}
\label{Lambda-reach-semi-dis1}
\end{figure}
We can take the signal significance as 
$S \equiv \frac{N_S}{\sqrt{N_S + N_B}} \sim \sqrt{N_S}$ for $N_S \gg N_B$, 
as is the case for us. Here $N_{S,B}$
is the number of events for the signal, background. 
In Fig.\ref{Lambda-reach-semi-dis1} we show for the 
$e e {\bar \mu} {\bar \mu}$ final state 
3$\sigma$ sensitivity plots in the $\Lambda - {\pounds}$ plane with $S$ = 3 for the various operators, where ${\pounds}$ is the total 
integrated luminosity in ${\rm fb}^{-1}$. The reach in the scale $\Lambda$ intially grows fast with ${\pounds}$ but more or less saturates at a value 
of ${\pounds}$ = 250 ${\rm fb}^{-1}$. One can see from the topmost curve (operators $S_1$ -- $S_6$) that the 3$\sigma$ sensitivity reach of the scale 
$\Lambda$ is 600 GeV for an integrated luminosity of 100 ${\rm fb}^{-1}$. Similar plots can also be obtained in this way for the 
$\mu \mu {\bar \tau} {\bar \tau}$ final state, which show slightly lower sensitivity reach. 

We wish to highlight an important point here. A value around 600 GeV for $\Lambda$ is only marginally bigger than $\sqrt{{\hat s}_{max}} \sim$ 420 GeV
for maximal parton luminosity in 14 TeV pp collisions, as mentioned above. This means that, if indeed four lepton flavor violation is discovered at
the LHC, the incompleteness of the effective Lagrangian approach is expected to show up and detailed features of the underlying new physics are likely 
to emerge.  

\section{Concluding Summary}
In this paper we have proposed a search for flavor violating quadrileptons without any hard jet or missing $E_T$ in the final state of 14 TeV pp
collisions at the LHC. We have motivated our proposal by an illustrative calculation of a significant cross section for producing the 
$ee {\bar \mu} {\bar \mu}$ (or ${\bar e} {\bar e} \mu \mu$) final state in an ${\rpv\lfv}$MSSM model. We have backed this up by a general analysis
of all possible six fermionic operators with SM fields in the chiral basis that contribute to the relevant partonic subprocesses. We have then 
computed the concerned cross section for each operator in 14 TeV pp collisions, assuming single operator dominance. We have also discussed the
sensitivity reach for the cutoff $\Lambda$ in terms of the integrated luminosity $\pounds$ at the LHC. We strongly urge our experimental colleagues
to undertake searches for such final states. Lepton flavor conservation may not hold at the terascale. 

\section*{ACKNOWLEDGMENTS}
We have gained from helpful conversations with S. Banerjee, A. Cooper-Sarkar, A. Datta, H. B. Dreiner, H. E. Haber, M. Hirsch, A. Kundu, W. Porod and 
S. Sarkar. We thank Pradipta Ghosh for technical help. DKG acknowledges partial 
support from the Department of Science and Technology, India under the 
grant SR/S2/HEP-12/2006. PR would like to 
acknowledge the hospitality of the Hans Bethe Centre of the University 
of Bonn and of the Instituto de Fisica Corpuscular of the University of Valencia. He has been supported in part by a DAE Raja Ramanna Fellowship.

\appendix


\end{document}